\documentclass[11pt,twoside]{article}
\usepackage{asp2010}

\resetcounters

\defcitealias{8}{PZ11}
\def\Msun{$M_\odot$}
\def\msun{$M_\odot$}

\begin{document}

\title{Radiation hydrodynamics of core-collapse supernovae: the ``key'' asset for a 
self-consistent modelling of these events}
\author{M.~L. Pumo$^{1,2,3}$ and  L. Zampieri$^1$
\affil{$^1$INAF-Osservatorio Astronomico di Padova, Vicolo dell'Osservatorio 5, I-35122 Padova, Italy}
\affil{$^2$Bonino-Pulejo Foundation, Via Uberto Bonino 15/C, I-98124 Messina, Italy}
\affil{$^3$INAF-Osservatorio Astrofisico di Catania, via S. Sofia 78, I-95123 Catania, Italy}}

\begin{abstract}
We have developed a specifically tailored relativistic, radiation-hy\-dro\-dyna\-mics Lagrangian 
code, that enables us to simulate the evolution of the main observables (light curve, 
evolution of photospheric velocity and temperature) in core-collapse supernova (CC-SN) events. 
The code features, some simulations as well as the implications of our results in connection with 
a possible ``standardization'' of the hydrogen-rich CC-SNe are briefly discussed. 
The possible role of this code in the development of a ``CC-SNe Laboratory'' for describing 
the evolution of a CC-SN event in a ``self-consistent'' way (from the main sequence up to the 
post-explosive phases) from a model/data comparison of light curves and spectra is also addressed.
\end{abstract}

%____________________________________
\section{Introduction}
Core-collapse supernova (CC-SN) events are thought to be the final explosive evolutionary phase of 
stars having initial (i.e. at main sequence) mass larger than $\sim$ 8-10 \msun\, \citep[e.g.][]{3}. 

Despite the importance of these explosive events in astrophysics, there are still basic questions to 
be answered (for details see e.g.~\citealt{8} --- PZ11 hereafter --- and references therein), related 
to the extreme variety of CC-SNe displays and linked to the uncertainties in the modelling of stellar 
evolution and explosion mechanism \citep[see e.g.][and references therein]{9}. In particular the exact 
link between the physical properties of the explosion (ejected mass, explosion energy, stellar structure 
and composition at the explosion) and the observational characteristics is far from being well-established, 
and a ``self-consistent'' description of CC-SN events (from the quiescent evolutionary phases up the 
post-explosion evolution) is still missing.

In this context, the creation of specific modelling tools that link progenitor evolution and explosion 
models to the main observables (i.e. light curve, evolution of photospheric velocity and temperature) of 
CC-SNe is of primary importance for clarifying the nature of the CC-SN events. The development of the 
relativistic, radiation hydrodynamics code described here represents a key step in this direction. 

%____________________________________
\section{Code description and future developments}
The distinctive features of the code (described in detail in \citealt*{10}, and \citetalias{8}) are an 
accurate treatment of radiative transfer coupled to relativistic hydrodynamics, a self-consistent treatment 
of the evolution of the innermost ejecta taking into account the gravitational effects of the central compact 
remnant, and a fully implicit Lagrangian approach to the solution of the coupled non-linear finite difference 
system of relativistic radiation-hydro equations. With this code it is then possible to follow the fall back 
of material on the central remnant in a fully relativistic formalism.

Using this code, we computed a grid of 22 post-explosive models of CC-SN events. This grid enabled us to (a) 
study the role of the ``main'' parameters affecting the post-explosion evolution  of the CC-SN events (namely 
the ejected mass, the progenitor radius, the explosion energy, and the amount of $^{56}$Ni), (b) understand 
the physical origin of some correlations between the photometric and spectroscopic properties of hydrogen-rich 
CC-SNe, and (c) perform a preliminary analysis on their utilization for cosmological purposes. 

All the results are fully described in \citetalias{8}. Here we recall that we are able to reproduce the main 
features (peak luminosity and phase at maximum) of the light curve of SN 1987A with models having initial 
radius $R_0 = 3\times10^{12}$ cm, total initial energy $E=1$ foe, amount of $^{56}$Ni $M_{Ni}=0.07$\Msun, and 
envelope mass ranging between 16 and 18 \Msun. The luminosity in the radioactive tail predicted by the model 
is lower than the observed one. This is a consequence of fallback occurring during the evolution. We found that 
the innermost $\sim0.01$\Msun\, of the envelope, containing $\sim 2.4\times10^{-3}$\Msun\, of $^{56}$Ni, have 
been accreted onto the central remnant. The time evolution of the photospheric velocity and temperature of 
SN 1987A is also well reproduced.

We are working at present to check the validity of the results connected with the previous items (b) and (c) 
against a more extended grid of models that is being computed from realistic initial conditions. The aforementioned 
grid of models will also serve to build an extended database to be compared with observations of single SNe in 
order to infer their physical properties, in analogy to what already being done using models with simplified 
initial conditions (e.g.~SN 2007od; see \citealt{4}). Our medium- and long-term goal is the development of a 
sort of ``CC-SNe Laboratory'' in which our code is interfaced, in input, with other codes dealing with the calculations 
of the pre-SN evolution and explosive nucleosynthesis, and in output with spectral synthesis codes 
\citepalias[for details see][and references therein]{8}. This will allow us to describe the evolution of a CC-SN 
event in a self-consistent way as a function of initial mass, metallicity, stellar rotation, and mass loss history 
of the CC-SN progenitor.

%____________________________________
\acknowledgements M.L.P. acknowledges financial support from the Bonino-Pulejo Foundation.

%____________________________________
%\bibliography{aspauthor}

\end{document}